\documentclass{PoS}
\usepackage{url}
\usepackage{lineno}
\usepackage{float}
\usepackage{subcaption}

\title{Testing a Novel Self-Assembling Data Paradigm in the Context of IACT Data}

\ShortTitle{Testing a Novel Self-Assembling Data Paradigm}

\author{\speaker{A. Weinstein}$^{a}$, L. Fortson$^{b}$, T. Brantseg$^{a}$,  C. Rulten$^{b}$, R. Lutz$^{c}$, J. Haupt$^{d}$,
M. Kakhodaie Elyaderani$^{d}$, J. Quinn$^{e}$ \\
        E-mail: \email{amandajw@iastate.edu}

{\footnotesize
$^{a}$ Department of Physics and Astronomy, Iowa State University, USA;
$^{b}$ Department of Physics \& Astronomy, University of Minnesota, USA;
$^{c}$ Department of Computer Science, Iowa State University, USA;
$^{d}$ Department of Electrical and Computer Engineering, University of Minnesota, USA;
$^{e}$ School of Physics, University College Dublin, Ireland;
}}

\abstract{

  The process of gathering and associating data from multiple sensors or sub-detectors due to a common physical event (the process of event-building) is used in many fields, including high-energy physics and $\gamma$-ray astronomy. Fault tolerance in event-building is a challenging problem that increases in difficulty with higher data throughput rates and increasing numbers of sub-detectors. We draw on biological self-assembly models in the development of a novel event-building paradigm that treats each packet of data from an individual sensor or sub-detector as if it were a molecule in solution. Just as molecules are capable of forming chemical bonds, ``bonds'' can be defined between data packets using metadata-based discriminants. A database -- which plays the role of a beaker of solution -- continually selects pairs of assemblies at random to test for bonds, which allows single packets and small assemblies to aggregate into larger assemblies.  During this process higher-quality associations supersede spurious ones. The database thereby becomes fluid, dynamic, and self-correcting rather than static.  We will describe tests of the self-assembly paradigm using our first fluid database prototype and data from the VERITAS $\gamma$-ray telescope.

}

\FullConference{The 34th International Cosmic Ray Conference,\\
  30 July- 6 August, 2015\\
  The Hague, The Netherlands}

\begin{document}

\section{Introduction}
\label{sec:introduction}


Modern scientific research frequently produces vast amounts of data to package and process.
In high-energy physics and astrophysics data analysis is typically performed on two to three
different time scales: \textit{real-time} analysis performed immediately, \textit{prompt} analysis performed after several hours or days, and \emph{offline} analysis and reduction performed days or years later.
Real-time analysis requires \emph{event-building}---the association of data packets from multiple detector components into a coherent description of a physical event---to also occur in real time.
This constraint competes with the processing time required to detect and compensate for event-building errors arising from
hardware or software malfunctions or data bottlenecks.
For large volumes of data rebuilding events is impractical and faulty event associations become frozen into the archive.

We present here a \emph{fluid database} designed to implement \emph{fault tolerance} in event-building.
The need for real-time analysis (where speed is prioritized over accuracy) is accommodated by permitting associations in the database to be tapped at an early stage.
We also consider a prototype implementation developed for a test-case from the VERITAS $\gamma$-ray observatory.

\section{Fluid Database Concept}
\label{sec:fluid-datab-conc}

The fluid database concept draws on the idea of \emph{self-assembly}, in which structures (\emph{assemblies}) are engineered via the random association of molecules in solution.
\emph{Algorithmic self-assembly} models this process by a pool of abstract objects with simple bonding rules and repeated random draws of pairs of assemblies from that pool \cite{Winfree1998}. Assemblies may be \emph{complete} or \emph{partial}, with the smallest partial assembly being a single unbonded object. We extrapolate the abstraction further by noting that these abstract objects can represent data packets rather than molecules in solution, with metadata-based discriminants replacing a simplified model of chemical bonds. The fluid database---which plays the role of a beaker of solution---continually selects pairs of assemblies at random to test for bonds, allowing single objects and small assemblies to aggregate into larger assemblies.

The self-assembly paradigm provides a natural, instrument-agnostic way of implementing \emph{fault-tolerance}.
\begin{figure}[h]
  \centering
  \includegraphics[width=0.85\textwidth]{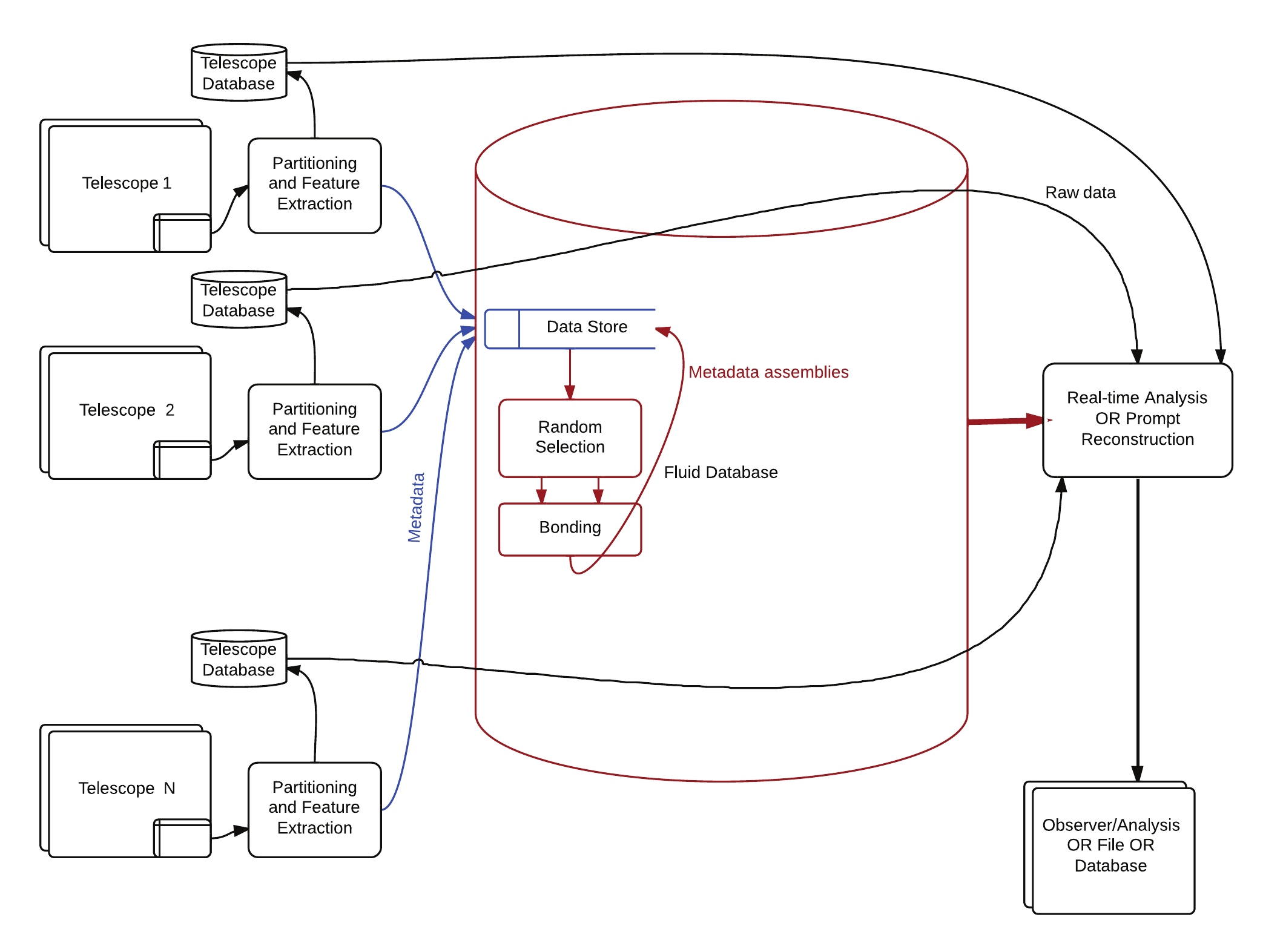}
  \caption{\small General flow of data within the self-assembly framework. Black arrows show raw data flow, while red arrows show metadata flow.}
  \label{fig:proposal-schematic}
\end{figure}
Fault tolerance typically consists of \emph{fault detection}, \emph{fault identification}, and \emph{fault recovery}. In conventional event-building, \emph{fault-identification} initially takes place after the fact, when problems with the data are recognized during data analysis or when the event-building process crashes due to some more egregious error. Custom \emph{fault detection} and \emph{recovery} code is then introduced into the event-building process in the form of special exceptions that recognize and address the faults. This procedure does nothing to prevent the initial failures and leads to a laundry list of exceptions embedded in the code. The fluid database, on the other hand, organically combines \emph{fault identification, detection,} and \emph{recovery} for a broad range of potential faults. The key is to correctly identify the metadata useful in associating the components of an event and to define the bonding criteria to allow for a range of bond strengths. In this way, two data packets may associate even if their metadata is not a perfect match. While some of these associations will be spurious, with enough iterations the fluid database is self-correcting, with the higher-quality associations superseding the spurious ones and data packets orphaned due to initially corrupt metadata being subsequently recovered.

\subsection{Tile Self-Assembly}
\label{sec:tile-self-assembly}

As previously noted, we draw heavily on the concept of algorithmic self-assembly \cite{Winfree1998} in developing the fluid database concept, and in particular on \emph{tile self-assembly} models, which abstract molecules as square tiles interacting in a two-dimensional space. These tiles may translate freely without rotating. The tiles are given ``glues'' on each side which are distinguished by means of colors or labels; the glues are further distinguished by assigning glues a numerical strength. Two tiles will interact if two sides with matching glues and sufficient strengths touch. In the simplest self-assembly model, the \emph{abstract tile assembly model} (aTAM), the assembly is given an ambient ``temperature'' value. Bonds with strengths lower than this value are not allowed to form, while bonds are permanent once formed. More sophisticated models introduce the concept of tiles detaching and reattaching from assemblies, allowing provisional bonds to be formed pending a better match for that tile, and/or use global information about the geometry of the assembly. These types of self-assembly allow for fault-tolerance to become a natural consequence of the assembly process itself \cite{DotyPatitz2010}. While the prototype fluid database described uses a more complex bond definition and takes some liberties with the simple square tile geometry, the tile self-assembly models nonetheless provide a valuable foundation for this work.

\subsection{Fluid Database in Context}
The fluid database is a random access database coupled to a continuously iterated process driven by a good pseudo-random number generator.
At each iteration two partial assemblies are randomly drawn and tested to see if they bond with one another and the database updated with the resulting assembly(ies).
Figure~\ref{fig:proposal-schematic} places this in context of the entire data flow diagram for a 'generic' telescope array. We consider here a use case where we partition the data into raw information and a \emph{metadata} component, with the raw information being stored in a separate telescope-specific database. When a telescope data packet is stored, the specific quantities and features are extracted, creating a much smaller \emph{metadata packet} that can be transmitted to another location, aggregated into event objects, and stored in an event database. The metadata packet also contains an identifier that points to a raw data packet in the telescope database. When analysis software (real-time or prompt) accesses an entry in the association database, the individual metadata packets within the association pull the raw data entries from the telescope databases to provide a full event.
This particular implementation reduces the storage and bandwidth requirements of the core self-assembly process.

\section{The VERITAS Test Case}
\label{sec:veritas-test-case}


The VERITAS array consists of four imaging atmospheric Cherenkov telescopes (IACTs), each of which consists of 496 individually triggered pixels. Location and timing information from the pixel-level (L1) triggers for each telescope is used to determine the telescope level (L2) trigger. The central array (L3) trigger requires at least two L2 triggers consistent with the arrival of Cherenkov light from the same shower in order to record an event.  The L3 trigger sends to each telescope a 48-bit serial stream containing the 32-bit event number, an 8-bit event type code, and an 8-bit `trigger mask' \cite{Weinstein2008}. The data recorded at each telescope is tagged by this 48-bit event mask as well as a timestamp provided by a local GPS clock \cite{HolderAtkins2006}. The raw data from all four telescopes and the array trigger is shipped to an event-building process (Harvester) in five bunched streams with a backup copy simultaneously recorded to disk. Ideally, event-building would be a fairly straightforward matter of matching event numbers, with the other quantities providing cross-checks on the primary association.


The most common event-building failure mode in VERITAS arises from bitwise corruption in the assigned event number during serial transmission.
The most usual effect is that data from one telescope is orphaned or attached to the wrong event.
Corruption of other parts of the event mask can also cause problems for the Harvester's error-checking code.
The timestamps themselves have been known to develop problems, most notably variations due to thermal drift.
These failure modes and their interactions make VERITAS an excellent small-scale test case for the fluid database paradigm.

\subsection{Assembling VERITAS data}
\label{sec:assembl-verit-data}


A nominal, complete VERITAS shower event will consist of one raw data packet recorded by each triggered telescope (a \emph{telescope event}), plus an additional data packet recorded by the L3 trigger (an \emph{L3 event}). Ideally, assembling the shower event requires that between two and four telescope events all connect to each other and the L3 event. However, this proves to be difficult to implement with the aTAM square-tile geometry.

We found the most obvious extensions of the square tile visualization scheme (such as allowing the tiles to assemble in three rather than two dimensions) to be flawed and overly complex. These schemes also generalize poorly to the $n$-telescope case.
The best option was an alternative visualization scheme dubbed the \emph{tinker-toy} model. Here the square tiles are replaced by individual points in a two-dimensional plane, with an additional central point representing the array (L3) trigger. (The location and spacing of these points need not correspond to the actual physical locations and spacing of the telescopes in the array.) The bonds between the tiles are represented by line segments connecting the points representing the individual tiles. Figure \ref{fig:assembly_concept} uses this visualization to depict the assembly process for a hypothetical assembly.

The tinker-toy model offers a more straightforward, easily generalized visualization of the self-assembly process.
It accommodates an arbitrary number of bonds between assembly components and thus generalizes easily into the $n$-telescope case.  It also demonstrably accommodates heterogeneous data (i.e. packets from more than one source).
The tinker-toy model also allows for a natural implementation of the pre-seeding feature of the fluid database, discussed in section \ref{sec:codebase}.
Finally, the geometry of the assembly carries important information.  As shown in Figure \ref{fig:assembly_concept}, assemblies for which the primary metadatum is correct in all cases are planar.  Assemblies that contain a recovered packet with corrupt metadata will be three-dimensional.

\section{Prototype Implementation}
\label{sec:prot-impl}

\subsection{Fluid Database}
\label{sec:fluid-database}

\begin{figure}[h]
  \centering
  \includegraphics[scale=0.4]{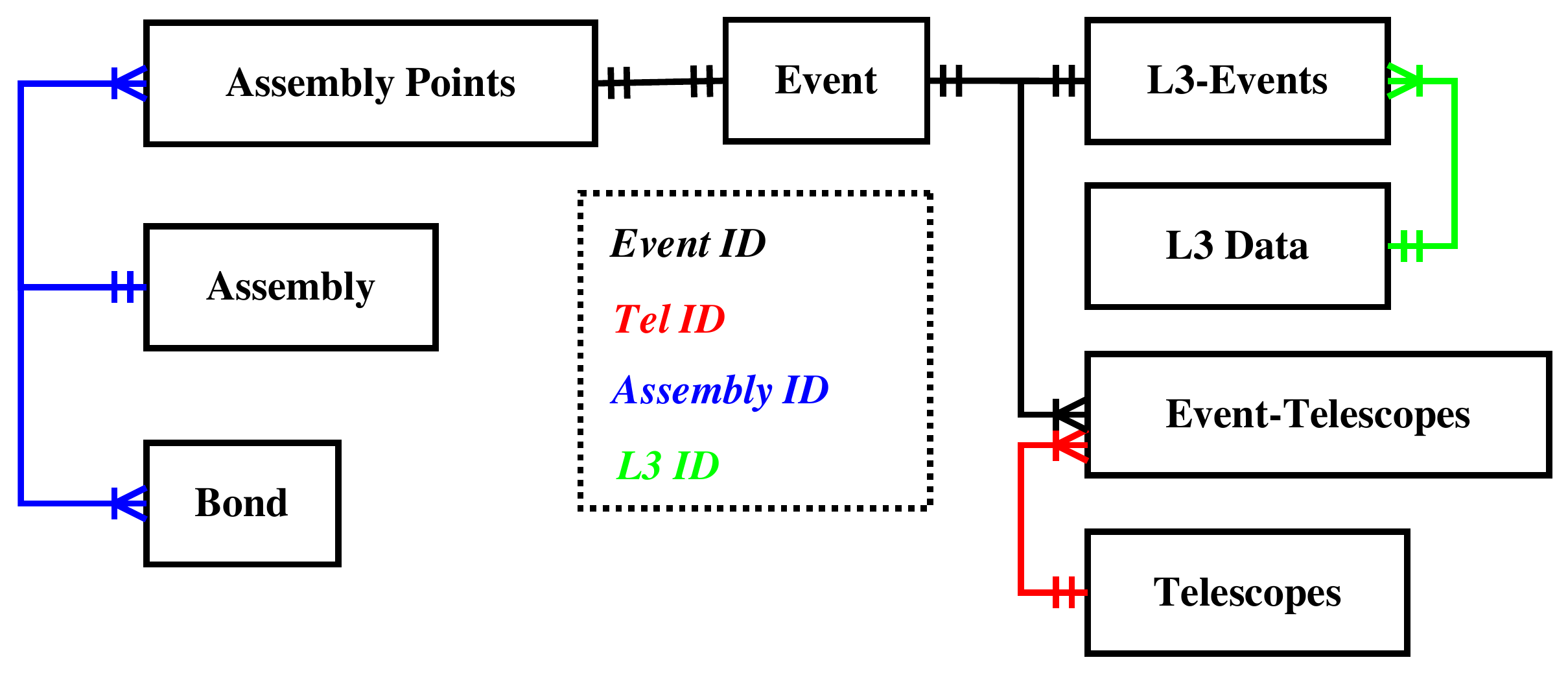}
  \caption{\small A simplified diagram of table relations within the PostgreSQL database. The connecting lines indicate primary key (two lines) or foreign key (arrow) constraints, with the key in question indicated by color as shown. 
  }
  \label{fig:DB_diagram}
\end{figure}

\begin{figure}[htbp!]
    \begin{subfigure}[t]{0.3\textwidth}
        \includegraphics[width=\textwidth]{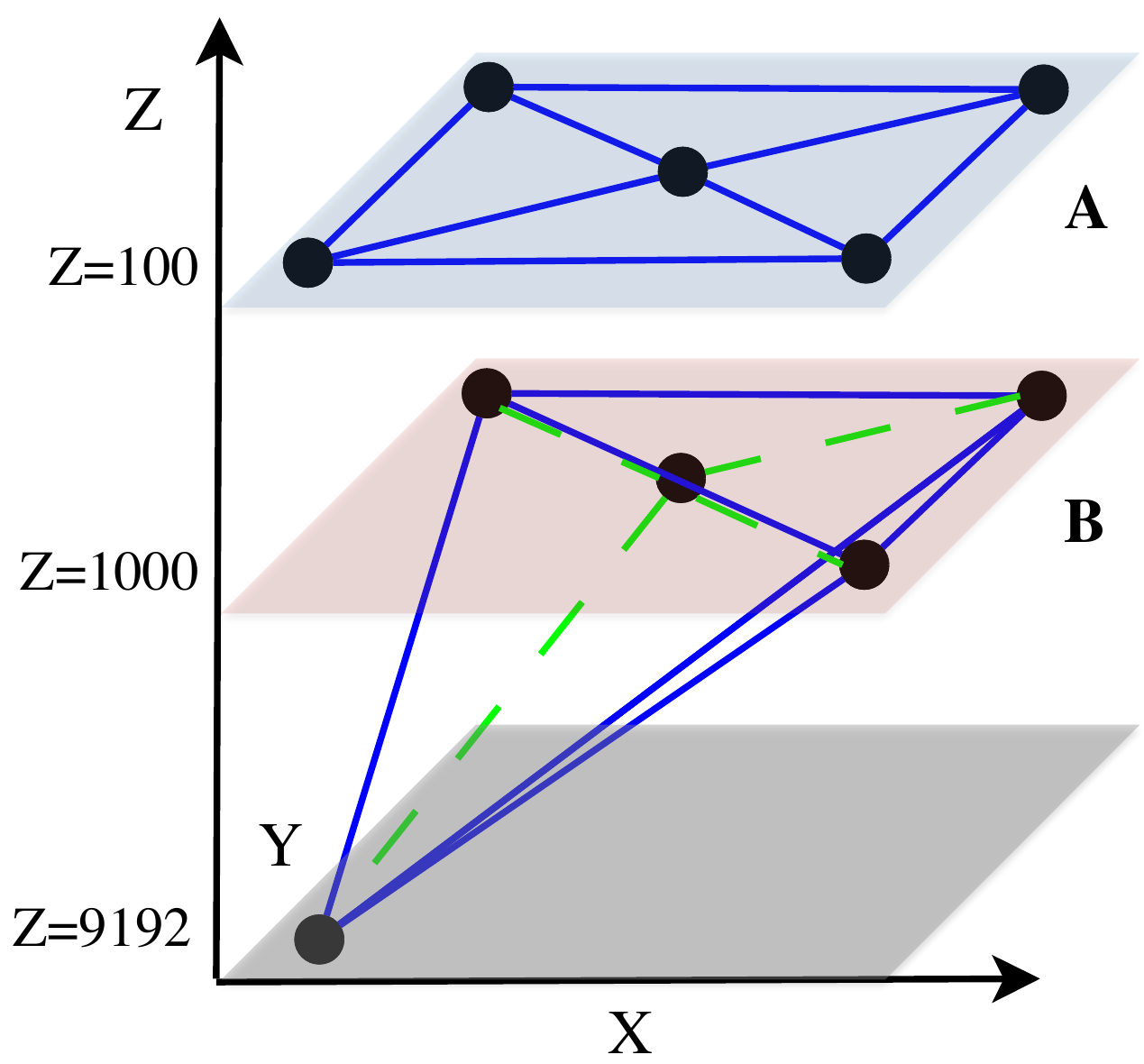}
        \label{fig:side_view}
    \end{subfigure}
   \begin{subfigure}[t]{0.6\textwidth}
        \includegraphics[width=\textwidth]{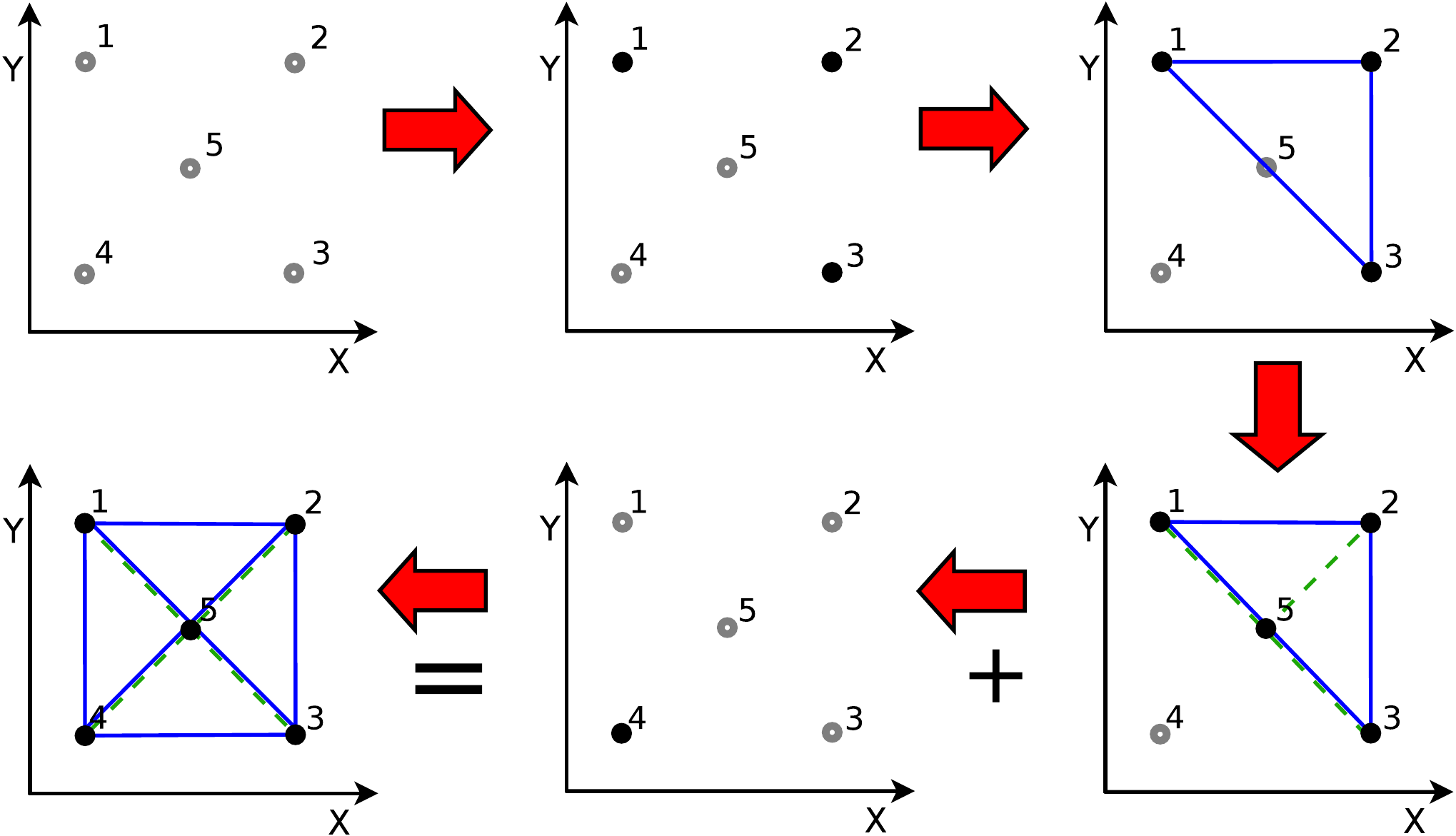}
        \label{fig:sample_cycle}
    \end{subfigure}
\caption{\small  Left: Side-on view of two assemblies in the spatial database for the VERITAS test case, using the tinker-toy model. X and Y coordinates in the spatial database identify specific data packets; Z identifies the event number assigned each packet by the L3 trigger.  Assembly A, which is perfect, lies purely in the z=100 plane as a result of pre-seeding with no errors in the primary metadatum. Assembly B does not. Right: Sample self-assembly cycle for Assembly B, assuming late arrival of the L3 data packets and single-bit corruption of the event number recorded by one telescope.  Clockwise from upper left:  top-down projection of spatial database, showing the loci for the four telescope packets and the L3 packet (point 5); a preliminary assembly at z=1000 (pre-seeded, no bonds) with 3 packets (filled black circles) at points 1-3;  the partial assembly with telescope-telescope bonds (solid blue); assembly with the L3 packet attached (dashed green lines); another partial assembly at z=9192 with one packet at point 4 combines with the previous assembly to form the final assembly, based on other metadata such as timestamp.}\label{fig:assembly_concept}
\end{figure}

In our prototype implementation, the fluid database is built using PostgreSQL 9.4 \cite{psql}, with the PostGIS 2.1 extension \cite{postgis}. This protocol allows us to treat the database both as a relational database, where objects are linked by shared pairs of keys, and a spatial database, where objects can be given a spatial location within the database. The addition of spatial properties to the data in the tinker-toy model, combined with the tinker-toy architecture shown in Figure \ref{fig:assembly_concept}, allows spatial queries to be used to check for pre-seeded assemblies prior to bonding and is also useful in mining the database for quality statistics. Figure \ref{fig:DB_diagram} shows how the relationships between tables are managed by key relations to ensure that stored data can be retrieved quickly and with a minimum of overhead.

The raw data making up each packet recorded by the VERITAS telescopes is stored in an \texttt{event} table. Each event entered into this table is automatically assigned a unique ID number, which is used to associate it to the telescope that recorded the event via the \texttt{event\_telescopes} table. This table, in turn, assigns a spatial location to the event via the telescope's entry in the \texttt{telescope} table. The event ID is also used to associate the event as a member of a specific assembly, as recorded in the \texttt{assembly} table. Each assembly is also assigned a unique ID, and this assembly ID is used to associate the assembly with the tile bonds that form it in the \texttt{bond} table. This interconnectivity of tables, shown in Figure \ref{fig:DB_diagram} ensures that the event table itself, which stores the raw data, is kept safe and unaltered by the assembly process, and also ensures that the final assemblies may be related back to their constituent events.

Modifications to the fluid database are managed by an external daemon process. This process controls the database via SQL calls issued to the database using the \texttt{libpqxx} 4.0.1 \cite{libpqxx} C++ API for PostgreSQL. The daemon continually loops through the following functions: retrieval of the state of assemblies stored in the database (including checking for recently-added data) and random selection of an assembly pair. This pair is evaluated by bonding code, which updates the database with the result. The continual monitoring of the state of the fluid database by the daemon allows checks to be made on the progress of the assembly process and allows for users to ``peek'' in for real-time or short-time data monitoring.




While the daemon is set up to mimic a Brownian bath process (molecules in solution meeting and interacting) we find that we can achieve significantly faster time-to-convergence by introducing three departures from pure randomness.
First, we \emph{parallelize} the random selection-bonding-update portion of the cycle.
Our prototype implementation is installed on a multi-node cluster, which enables us to do this using a message passing interface library such as OpenMPI, and by using a compatible parallel random number generator, in this case the Random123 library \cite{Salmon:2011:PRN:2063384.2063405}.
We plan to keep the database coherent by using a very simple statement-based replication approach. Here replication is defined to be the process of sharing transactional data to ensure consistency between redundant database nodes.
When an assembly is retrieved by the code for bonding, the daemon places a temporary lock on the assembly ID to mark that the assembly is \emph{checked out}, preventing potential errors caused by multiple processes attempting to modify an assembly at the same time.
When the bonding calculation is complete, the assemblies are \emph{checked in} and the temporary lock is released.
Second, we \emph{pre-seed} the database by adding data to the database as part of preliminary assemblies based on a single \emph{primary criterion} (in this case an exact match of event number).  Each preliminary assembly has its full bond strengths evaluated at the time of storage.
The diagram in Figure \ref{fig:assembly_concept} demonstrates how a pre-seeded preliminary assembly can interact with other assemblies in the database to produce final assemblies.
%
Pre-seeding is advantageous as long as there is a significant subset of \emph{perfect} (all bonds at maximum strength and presumed correct) \emph{complete} (all expected metadata packets present) assemblies that can be created on the basis of the primary criterion alone.
A sample of raw data from three VERITAS runs indicates that this is the case: rates of corrupted event numbers in raw data from three VERITAS runs were highly variable but ranged from 50-700 incorrect event numbers per run out of several hundred thousand total events.
Third, we permit partial or complete \emph{precipitation} of assemblies from the fluid database by reducing the random draw probability in inverse proportion to the \emph{completeness} and \emph{perfection} of the assembly.
In terms of our fluid metaphor, this can be thought of as molecules formed in a solution precipitating out to the bottom of the beaker.
In terms of practical implementation, we add a weighting function to the random number generator.
In the test case currently under study, completeness of the assembly is binary, since data from all telescopes and L3 should be recorded for every event.
For any assembly that is both complete and perfect the probability of drawing that assembly is dropped to zero; convergence may be further adjusted by tuning the probability of drawing complete but imperfect assemblies.
In a test case where completeness is not deterministic (i.e. only data from telescopes with L2 triggers is read out) probability of drawing an assembly would be suppressed (rather than zeroed out) in proportion to an estimated \emph{completeness probability}.  This type of test case is easily simulated using the available VERITAS data. Algorithms for estimating the completeness probability are under investigation.



\subsection{Bonding Algorithm and Framework}
\label{sec:codebase}
A set of C++ classes, implementing metadata packets, bonds, and assemblies provides a general and flexible framework for implementing the bonding algorithm.  The chief function of this code is to manage the creation, destruction, and definition of bonds, whose strength is calculated as a real-valued function of the relevant metadata of the two bonded packets as well as global evaluations of an assembly's quality.
Weaker bonds can be replaced within an assembly by stronger bonds from a newly merged assembly.

%
At the moment we are testing simple bond definitions based on low-level metadata (e.g. event numbers, GPS timestamps), but work is being done in parallel to include features extracted from the bulk data from the VERITAS cameras. Such criteria would add robustness to the bond definition in cases where the low-level metadata are corrupt. Our focus in this area is on two classes of features. The first general class is features with relations to physically meaningful data, such the energy or position of the incident $\gamma$-ray. For this class of features, we are evaluating feature extraction methods ranging from classical linear prediction models to more advanced nonlinear tools such as neural networks.

The second general class is motivated by analysis of intrinsic structure in the data, as well as characteristics of the background and ambient noise. For this class of features, we are using array-level simulations to identify these characteristics and develop a priority tree to determine when and where these features can provide supplemental information on the bond strength.

\section{Conclusions}
\label{sec:conclusions}
We present here a preliminary, in-development prototype of the \emph{fluid database}. We are in the process of establishing its key performance benchmarks.  Chief among these is the dependence of the fraction of correctly-formed complete assemblies on the number of input events, the number of flawed inputs, and elapsed time. Benchmarks are subject to the additional constraint that the fluid database provide preliminary assemblies to the real-time analysis at a speed comparable to that of the Harvester.
Extensions of this approach to the generic \emph{n}-telescope case have implications for future Cherenkov observatories in particular and large sensor arrays in general.

\section{Acknowledgements}
\label{sec:acknowledgements}

This research is supported by Award \# NSF/PHY-1419240 and  Award \#  NSF/PHY-1419259 from the National Science Foundation.
VERITAS is supported by grants from the U.S. Department of Energy Office of Science, the U.S. National Science Foundation and the Smithsonian Institution, and by NSERC in Canada. We acknowledge the excellent work of the technical support staff at the Fred Lawrence Whipple Observatory and at the collaborating institutions in the construction and operation of the instrument.

The VERITAS Collaboration is grateful to Trevor Weekes for his seminal contributions and leadership in the field of VHE gamma-ray astrophysics, which made this study possible.

\bibliographystyle{JHEP}
\bibliography{../biblio}

\end{document}